\documentclass[aps,prl,preprint,superscriptaddress,noshowpacs,titlepage,10pt]{revtex4}

\usepackage{CJK}
\usepackage{graphicx}
\usepackage{dcolumn}
\usepackage{bm}

\begin{document}
\begin{CJK*}{UTF8}{gbsn}
\preprint{}

\title{Sculpting the Vertex: Manipulating the Configuration Space Topography and Topology of Origami Vertices to Design Mechanical Robustness}

\author{Bin Liu}
\affiliation{Department of Physics, Cornell University, Ithaca, NY 14850}
\affiliation{School of Natural Sciences, University of California, Merced, Merced, CA 95343, USA}
\author{Arthur A. Evans}
\affiliation{Department of Physics, University of Massachusetts Amherst, Amherst, Massachusetts 01003, USA}
\author{Jesse L. Silverberg}
\affiliation{Department of Physics, Cornell University, Ithaca, NY 14850}
\author{Christian D. Santangelo}
\affiliation{Department of Physics, University of Massachusetts Amherst, Amherst, Massachusetts 01003, USA}
\author{Robert J. Lang}
\affiliation{Lang Origami, Alamo, California 94507}
\author{Thomas C. Hull}
\affiliation{Department of Mathematics, Western New England University, Springfield, MA 01119, USA}
\author{Itai Cohen}
\affiliation{Department of Physics, Cornell University, Ithaca, NY 14850}
\date{\today}

\pacs{}

\keywords{}
\maketitle
\end{CJK*}


\keywords{}
\textbf{
The geometric, aesthetic, and mathematical elegance of origami is being recognized as a powerful pathway to self-assembly of micro and nano-scale machines with programmable mechanical properties \cite{Lv:2014cs, Filipov:2015cr, tactom:2016tu, Greenberg:2011bs, Hanna:2014fp, You:2007do, Onal:2011da, Overvelde:2016gn, Kuribayashi:2006gh, Dureisseix:2012tk, Yasuda:2015eg, Waitukaitis:2015dk, Evans:2015gu}. The typical approach to designing the mechanical response of an ideal origami machine is to include mechanisms \cite{Latombe:1991, Erdman:1997wi} where mechanical constraints transform applied forces into a desired motion along a narrow set of degrees of freedom \cite{Lv:2014cs, Filipov:2015cr, tactom:2016tu, Greenberg:2011bs, Hanna:2014fp}. In fact, to date, most design approaches focus on building up complex mechanisms from simple ones in ways that preserve each individual mechanism's degree of freedom (DOF), with examples ranging from simple robotic arms \cite{You:2007do, Onal:2011da, Overvelde:2016gn} to homogenous arrays of identical vertices \cite{Kuribayashi:2006gh, Dureisseix:2012tk, Yasuda:2015eg, Waitukaitis:2015dk, Evans:2015gu}, such as the well-known Miura-ori \cite{MIURA:1985tt}. However, such approaches typically require tight fabrication tolerances, and often suffer from parasitic compliance \cite{Tachi:2011tx, Francis:2014dp, Hanna:2014fp, Chen:2015sci}. In this work, we demonstrate a technique in which high-degree-of-freedom mechanisms associated with single vertices are heterogeneously combined so that the coupled phase spaces of neighboring vertices are pared down to a controlled range of motions. This approach has the advantage that it produces mechanisms that retain the DOF at each vertex, are robust against fabrication tolerances and parasitic compliance, but nevertheless effectively constrain the range of motion of the entire machine. We demonstrate the utility of this approach by mapping out the configuration space for the modified Miura-ori vertex of degree 6, and show that when strung together, their combined configuration spaces create mechanisms that isolate deformations, constrain the configuration topology of neighboring vertices, or lead to sequential bistable folding throughout the entire origami sheet.
}

\begin{figure*} \includegraphics[width=.8\textwidth]{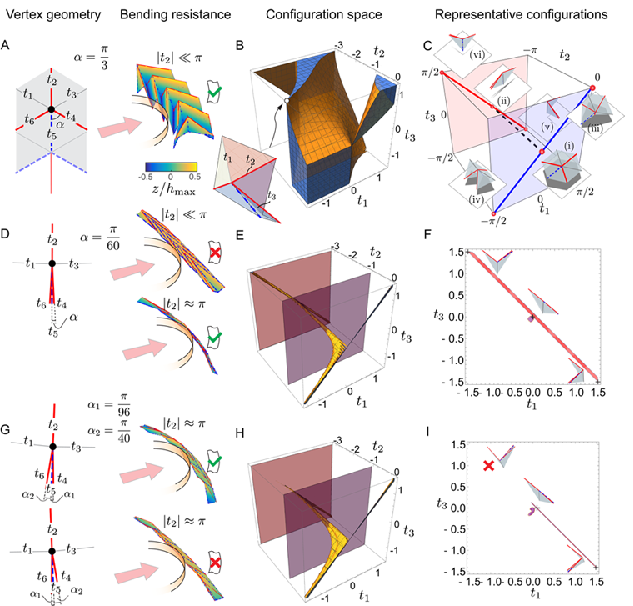}
\caption{Mechanics of a degree-6 Miura-ori vertex. Columns from left to right correspond to vertex geometries, bending resistances, configuration space of accessible structures, and representative configurations, respectively. (A) The geometry of a degree-6 Miura-ori unit cell (shaded area) is given by a set of 6 creases converging at a vertex (left panel). Solid red and dashed blue lines are the predetermined mountain and valley folds, respectively, while the thin gray lines are two universal creases that are allowed to fold in either direction. 
The crease angle $\alpha$ (here, $\alpha=\pi/3$) corresponds to the angle between the valley and mountain folds. Also shown is a column of tiled vertices with the color indicating the relative height. We find for this choice of crease angle, the column can accommodate a large in-plane bending angle without intersections between facets through asymmetric folding of the universal creases (right panel).
(B) The phase space consisting of the accessible configurations for this vertex can be parametrized by 3 folding angles, $t_1, t_2$, and $t_3$. A large fraction of the configuration space is accessible, including asymmetric folding geometries where $t_1\neq t_3$. The inset shows a schematic of an extreme asymmetric structure with $t_1=-t_3=-\pi/2$ and $t_2\approx -1$. 
(C) A skeletal representation of the configuration space along with red dots marking illustrative locations and schematics of their corresponding vertex structures. 
Here, structures along the blue line ($t_1=t_3, t_2=0$) are symmetric, while structures along the red line ($t_1=-t_3, t_2=-\pi$) are antisymmetric. The dashed line corresponds to the regular Miura-ori folding, where the universal creases remain unfolded, i.e., $t_1=t_3=0$. (D) A tiled column of vertices with $\alpha=\pi/60$ allows minimal in-plane bending until it is flat folded. The configuration space for this vertex (E) is significantly reduced. For example, a finite mountain fold $t_2=-1$ rad (purple plane in (E)), reduces the configuration space to a single spot (F), with $t_1 \approx t_3 \approx 0$. The structure thus folds as though the universal crease were rigid and flat. Once $t_2$ folds flat ($t_2 = -\pi$), the structure is allowed to deform anti-symmetrically with $t_1\approx -t_3$. Again we mark illustrative locations and schematics of their corresponding vertex structures. (G) For an asymmetric vertex, with two different crease angles $\alpha_1=\pi/96,\alpha_2=\pi/40$ (left panel), the allowed in-plane bending is asymmetric and the configuration space (H) is again significantly reduced. For example, a finite mountain fold $t_2=-1$ rad (purple plane in (H)), reduces the configuration to a comma shaped spot (I) with $t_1\approx t_3\approx 0$. However, once $t_2\approx -\pi$, only half of the antisymmetric $t_1\approx -t_3$ region ($t_1>0$ and $t_3<0$) is accessible, which gives rise to a diodic in-plane bending. The allowance and forbiddenness of all those structures to in-plane bending are indicated by the ``check'' and ``cross" symbols respectively. 
}\label{fig01}
\end{figure*}

The basic principles of an origami mechanism are nicely illustrated by the tessellation of the Miura-ori \cite{MIURA:1985tt}. The fold pattern for this vertex is illustrated by the dashed blue and solid red lines in Fig.~\ref{fig01}(A). Here, 4 creases converge at a single degree-4 vertex. To determine its degrees of freedom (DOF) it is necessary to account for the various constraints on the configurations. Holding a degree $N$ vertex fixed at the origin, each freely rotating crease contributes $2N$ degrees of freedom. From these DOF we subtract the $N$ constraints due to the angles between the creases in the fold pattern, and 3 rotational symmetries. Thus using the degree counting rule DOF=$2N-N-3$ gives a single degree of freedom for each degree 4 Miura-ori vertex \cite{Hull:2006, Evans:2015et}. Moreover, a tessellation of such vertices is synchronized and characterized by a single degree of freedom so that the folding of any individual crease controls the folding of the entire sheet \cite{Tachi:2009ub}. This mathematical argument however ignores real world aspects of sheets, including the bending of facets, finite sheet thickness, errors in folding angles, and even distortions of the crease lines \cite{Wheeler:2016kw}. Such non-ideal behaviors lead to parasitic compliance including out of plane bending and twisting \cite{Schenk:2013kk, Wei:2013kn}, as well as inhomogeneous compression and defects \cite{Silverberg:2014dn, Evans:2015et}. As such these imperfections often complicate and even destroy the single degree of freedom mechanism designed into this and other structures.  

Here we propose an orthogonal approach to creating mechanisms in origami structures. Rather than using the number of constraints to reduce the range of motion for each vertex, we employ the angles in the fold pattern to restrict the configuration space and guide vertex-vertex interactions to achieve a desired mechanism. We illustrate this approach using a degree-6 Miura-ori, where each vertex has $N=6$ creases with folding angles $t_1$, $t_2$, $\cdots$, $t_6$. In addition to the four predetermined mountain-valley folds in a typical Miura-ori (color lines in Fig.~\ref{fig01}(A)), two universal creases ($t_1$ and $t_3$) along the diagonal of the parallelograms adopt either mountain or valley folds (thin gray lines Fig.~\ref{fig01}(A)). Thus, each vertex has DOF$=12-6-3=3$ and can potentially accommodate a wide range of deformations \cite{Evans:2015et}. For example when the angle between the mountain and valley creases $\alpha=\pi/3$, a vertical tessellation of such vertices can include a large amount of in-plane bending in either direction due to the asymmetric folding of the two universal creases (Fig.~\ref{fig01}(A)). 

The configuration space of this degree 6 vertex can be visualized by distinguishing between allowed (filled) and forbidden (empty) regions parametrized by the three folding angles $t_1$, $t_2$, $t_3$ (Fig.~\ref{fig01}(B)) (See Methods). To illustrate the accessible structures, we dissect the 3D configuration space by marking 6 representative spots accompanied with the corresponding vertex folds (Fig~\ref{fig01}(C)): (i), the unfolded configuration ($t_1=t_2=t_3=0$); (ii), the flat folded configuration ($t_1=t_3=0$ and $t_2=-\pi$); (iii) and (iv), two symmetrically folded structures associated with out-of-plane deformations ($t_1=t_3=\pm \pi/2$ and $t_2=0$); (v) and (vi), two antisymmetric folded structures associated with in-plane bending ($t_1=-t_3=\pm \pi/2$ and $t_2=-\pi$). The black dashed line shows the path associated with the one degree of freedom folding of an ideal Miura-ori vertex ($t_2=t_3=0$). The blue line ($t_2=0$ and $t_1=t_3$) illustrates symmetric folding configurations. Finally, the red line ($t_2=-\pi$ and $t_1=-t_3$) corresponds to antisymmetric folding configurations for a flat-folded structure that is bent either to one side or the other.

Remarkably, decreasing the fold angle to $\alpha=\pi/60$, maintains the dimension of the accessible configuration space while substantially altering its range of deformations (Fig.~\ref{fig01}(D)). For example, a vertical tessellation of these vertices is highly resistant to in-plane bending until the sheet is completely flat folded with $t_2=-\pi$. The origin of this rigidity can be traced to the substantially restricted configuration space shown in Fig.~\ref{fig01}(E). This configuration space approaches the skeletal form represented by the blue, dashed black, and red lines in Fig.~\ref{fig01}(C). To emphasize the reduction in conformations we illustrate the areal projections of this skeletal configuration space onto two planes, one at $t_2=1$ (purple plane) and a second at $t_2\approx -\pi$ (red plane). As shown in Fig.~\ref{fig01}(F) the configuration space at $t_2=1$ is reduced to a single dot (purple region), thereby excluding structures such as that shown in the inset to Fig.~\ref{fig01}(B). In addition, we find that the only pathway available for asymmetric folding, is for $t_2\approx -\pi$ so that the structure is flat folded and the universal creases can fold anti-symmetrically with $t_1\approx-t_3$ (red linear region and accompanying folded structures). 

Breaking the reflection symmetry of the vertex by having two different angles $\alpha_1,\alpha_2$ between the mountain and valley creases creates diode like in-plane bending behavior. Similar to the case for symmetric but small $\alpha$,  a vertical tessellation of these vertices is highly resistant to in-plane bending until $t_2=-\pi$. However, even in this extreme case the structure only bends in one direction but not the other (Fig.~\ref{fig01}(G)). We find that the configuration space of such vertices is missing one of the antisymmetric branches (Fig.~\ref{fig01}(H,I)). This asymmetry in the configuration space in-turn gives rise to the diodic behavior with bending only towards the side with the larger crease angle. 

\begin{figure*} \includegraphics[width=.8\textwidth]{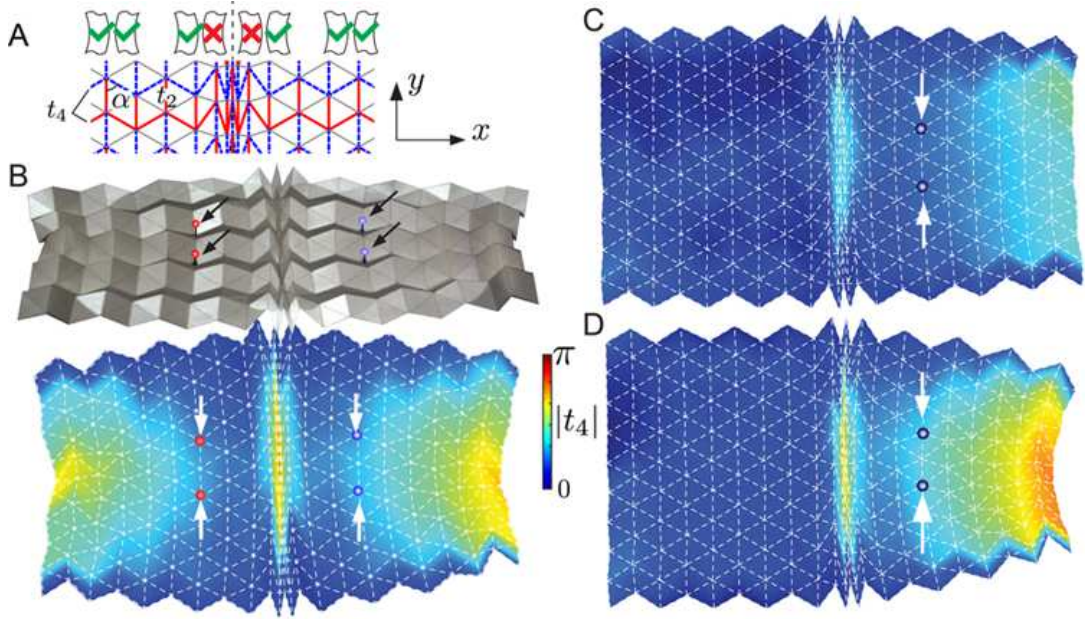}
\caption{Designed mechanical isolation. (A) A left right symmetric design where diodic columnar tilings (Fig.~1(G-I)) are sandwiched between symmetric columnar tilings. The outer symmetric vertices are flexible and can bend either way. The inner symmetric columnar tiling with small $\alpha$ is relatively rigid, but capable of bending in either direction once it is flat folded. Finally, the ``check'' and ``cross'' schematics illustrate whether in-plane bending toward the left or right is allowed at each diodic vertex. The particular sequence shown is designed to prevent in-plane bending on one half of the sheet from crossing over to the other half. (B-D) Experimental testing of the design in (A). (B) 
A 3D reconstruction of a real folded structure pinched at two locations (top) along with the extracted folding angle $t_4$ indicated by an interpolated color map (bottom). The sheet response indicates both sides can deform.     
(C-D) Under asymmetric pinching the double diode configuration isolates one side of the sheet from the other serving as a strain isolation barrier under both moderate (C) and extreme (D) deformation. 
}\label{fig02}
\end{figure*}

Our configuration space analysis immediately suggests a strategy for designing mechanical metamaterials that take advantage of these vertex properties to isolate in-plane bending deformations. To demonstrate this behavior we design an origami structure that combines two opposite diotic columnar tilings, and sandwiches them between symmetric degree-6 Miura-ori tilings (Fig.~\ref{fig02}(A)). This fold pattern is designed to prevent in-plane bending on one side of the sheet from affecting the deformations on the other side. To assess our design, we folded the origami structure, used pins to apply in plane deformations, imaged it, and rendered the three dimensional conformation of the structure and the pins as shown in Fig.~\ref{fig02}(B) (See Methods). To characterize the deformation at each vertex we keep track of the folding angle $t_4$, which is a sensitive indicator of the extent to which a given vertex is folded. We find that pinching due to the pin configuration induces large deformations on both sides of the sheet indicating that both halves are free to deform. When the pinching due to the pins is only applied to the right side, however, we find that the deformations are isolated and do not penetrate past the double diode structure, in accordance with the design criteria (Fig.~\ref{fig02}(C,D)). 

\begin{figure} \includegraphics[width=.7\textwidth]{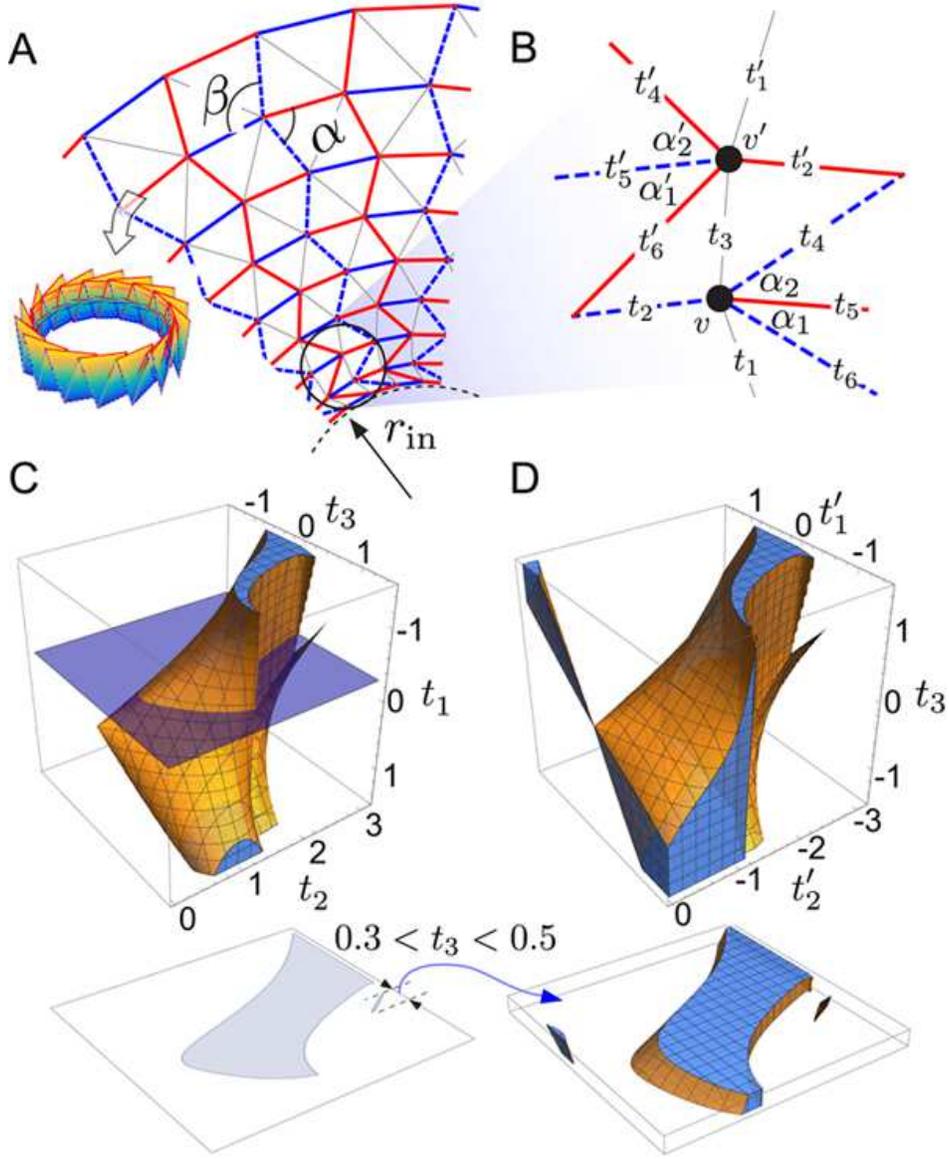}
\caption{Designed sequential bifurcations due to neighboring vertex interactions. (A) Flat-foldable vertices are tessellated in a circular washer geometry by requiring the sum of two opposite crease angles $\alpha+\beta=\pi$. The angle $\alpha$ vanishes at finite radius $r_{\mathrm {in}}$ which determines the size of the central hole when the sheet is fully compressed into a cylindrical wall (inset). (B) Two radially neighboring vertices labeled $v$ and $v'$ interact by sharing a common crease $t_3$. (C) The configuration space of vertex $v$ can be parametrized by 3 folding angles, $t_1, t_2$,and $t_3$. This configuration space has different topologies depending on the restrictions placed by folds shared with neighboring vertices and by $r_{\mathrm{in}}$. For example, for positive $r_\mathrm{in}$, $t_1$ is restricted to be less than zero (upper half of the configuration space). The constraint $t_1=-0.3$ (blue plane) yields a topologically disconnected configuration space (lower cross section) where the large region corresponds to a folded structure that is still flexible and the top right sliver with $0.3 < t_3 < 0.5$ corresponds to a highly constrained or locked structure when the vertex snaps into a fully folded state. 
(D) The configuration space of vertex $v'$ can be parametrized by 3 folding angles, $t_1'$, $t_2'$, and $t_3$.  This configuration space also has different topologies depending on the restrictions placed by the fold shared with vertex $v$.
The constraint $0.3 < t_3 < 0.5$ in (C) restricts the configuration space of $v'$ as shown in the lower image. 
Once again, the configuration space is topologically disconnected with the upper right sliver corresponding to a collapsed vertex. By collapsing $v'$, 
a similar disconnected topology is generated in the next vertex out. Thus, the sequential collapse can proceed across successive layers until the entire structure is flat-folded into a cylindrical wall. 
}\label{fig03}
\end{figure}

There are however a couple of disadvantages for this design. First, it is not flat foldable. Second, since strain due to in-plane bending is focused along the central axis, under even moderate strains the sheet as a whole becomes quite rigid and more susceptible to tearing as its size increases. Significantly greater in plane bending for arbitrarily large sheets can be accommodated in structures where the creases $t_2$ and $t_5$ are no longer co-linear. For example, the degree-6 Miura-ori washer geometry shown in Fig.~\ref{fig03}(A), one of the so called spiral origami structures \cite{Nojima:2002ws}, is a fully collapsible structure that forms a cylinder in its folded state. Here, the angle between the mountain and valley folds, $\alpha$, decreases with radius $r$ as $\alpha=\cos^{-1}(r_\mathrm{in}/r)-\pi/N_\phi$, where $r_\mathrm{in}$ is the inner radius of this structure and $N_\phi$ is the number of vertices along the angular direction. The corresponding opposite angles $\beta$ satisfies $\alpha+\beta=\pi$, as a necessary condition for flat-foldable degree-6 Miura-ori vertices. 

Relaxing the co-linearity constraint for creases $t_2$ and $t_5$ (Fig.~\ref{fig03}(B)) substantially alters the configuration space as shown in Fig.~\ref{fig03}(C)). In particular, we find that the configuration space has bifurcations as indicated by the distinct conical section oriented towards the upper right.  Importantly, this bifurcation can be used to topologically protect the flat folded configuration. By setting $r_\mathrm{in}>0$ and $N_\phi=32$, we can force $t_1=-0.3$ in the flat folded state. For this value of $t_1$ the configurations space is topologically disconnected (blue plane and lower image in Fig. \ref{fig03}(C)). For the topologically disconnected configuration subspace, the large area corresponds to a very compliant and deformable structure whereas the sliver between $0.3 < t_3 < 0.5$ corresponds to a highly constrained and collapsed structure. The narrowness of the configuration space for the collapsed structure suggests that it is rigid and capable of strain isolation. To jump or snap between these configurations, the sheet needs to access higher energy deformation modes such as stretching of the facets, which are not accounted for by the configuration space analysis \cite{Silverberg:2015gb}. 

The bifurcated configuration space also immediately suggests a mechanism in which successive layers of the structure can be made to sequentially snap onto the collapsed inner cylinder. Specifically, we choose the radial separation of the neighboring vertex so that the bifurcation in \emph{its configuration space} is present when the shared universal crease $0.3 < t_3 < 0.5$ corresponding to the topologically protected state for $v'$ (Fig.~\ref{fig03}(D)). This design ensures that once the inner vertex is locked into the highly constrained collapsed structure, the second layer will become bistable so that it can snap onto the first. It can be shown that by growing the distance between vertices in adjacent layers linearly with radius, this sequential snapping continues until the entire sheet collapses onto a fully folded cylindrical structure that is rigid and isolates strain by maintaining the geometry of its inner boundary. 

\begin{figure} \includegraphics[width=.8\textwidth]{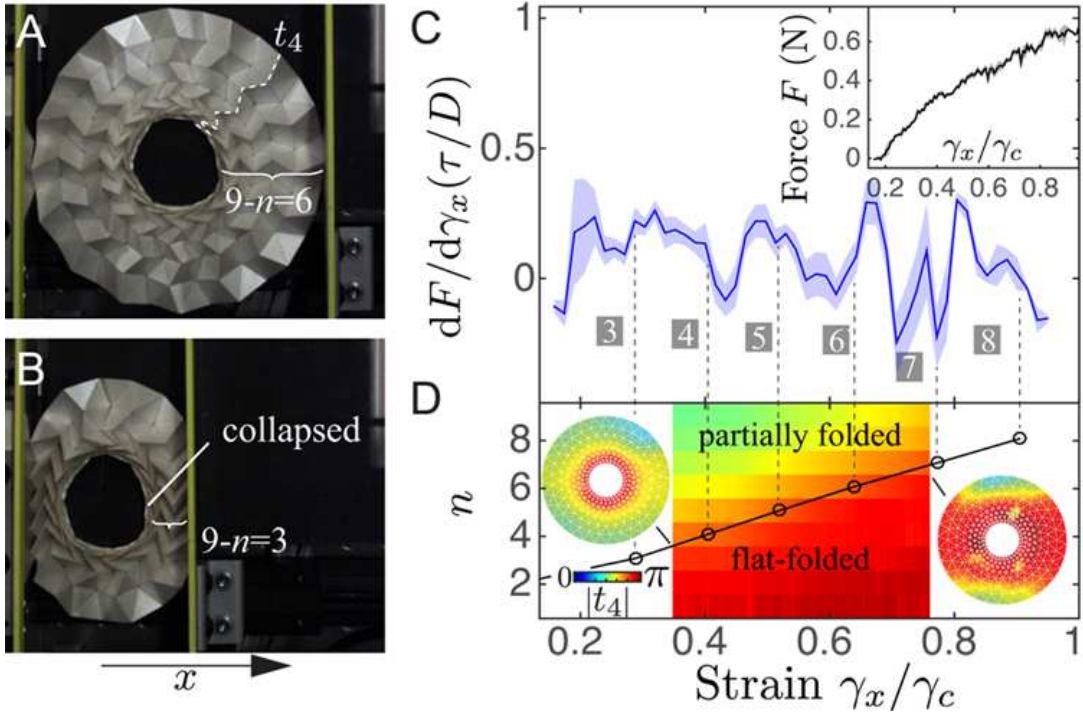}
\caption{Experimental measurement of designed sequential bifurcation mechanics. (A) and (B) The origami washer is deformed uni-axially between two parallel plates. The number $n$ shows the number of layers within a total of 9 with vertices that snapped into their fully folded configurations. (C) The compressive modulus of the structure $K\propto \frac{dF}{d\gamma_x}$ is obtained from the force-strain relationship $F$ vs. $\gamma_x/\gamma_c$. Here $\gamma_c=0.75$ and corresponds to the strain at which the entire sheet snaps into a cylindrical wall. The modulus $K=\frac{dF}{d\gamma_x} (\tau/D)$, normalized by flexural rigidity $D$ and thickness of the sheet $\tau$, shows a sequence of sharp drops in its value each of which corresponds to a snap. (D) The two circular insets indicate the full 2D distribution of the fold angle $t_4$ at the early and late portion of the experiments mapped onto the unfolded sheet. The central figure shows a color map indicating the averaged angle value of the $t_4$ fold for each layer (1-9) as the normalized strain $\gamma_x/\gamma_c$ is increased. We find that the structure folds from the inside out so that the inner layers have larger $t_4$ values than the outer layers. Finally, we simulated uniform compression of this structure and overlay the normalized strain values at which each layer undergoes a snapping transition (circles). The dashed lines are guides to the eye.  
}\label{fig04}
\end{figure}

To test these design principles, we fabricate a Miura-ori washer following the requirement for a vertex separation that grows linearly with radius. The structure is placed between two plates and compressed laterally (Fig.~\ref{fig04}(A,B)). We find that the structure is able to accommodate a large degree of compressive strain $\gamma_x = (l_0-l)/l_0$, where $l_0$ and $l$ correspond to the lateral size of the unfolded and folded sheet respectively. During compression, the innermost layer is the first to collapse. Upon further compression, successive layers collapse onto the inner layer (see SI video). At the same time the remaining uncollapsed layers retain their flexibility.  Importantly, throughout the entire folding process up to where the strain reaches $\gamma_c$ indicating full collapse, the inner boundary maintains its shape indicating effective strain isolation. 

The snapping behavior of successive layers can be quantified by the compressive modulus or derivative of the applied compressive force $F$ with respect to the strain $\gamma_x$ (Fig.~\ref{fig04}(C)). We find that each snap, indicated by the number of collapsed layers $n$, corresponds to a rapid decrease of $dF/d\gamma_x$ as the normalized strain $\gamma_x/\gamma_c$ is increased (Fig.~\ref{fig04}(C)). This decrease arises from the extra space that is released when a layer snaps into place, which relaxes the folds in the remaining layers and reduces their resistance to compression. 

To verify that snapping proceeds from the inner boundary outwards, we use 3D reconstructions to track the folding angle $t_4$ and map its value using a color map onto the unfolded sheet configurations (left and right sides of Fig.~\ref{fig04}(D)). The radially averaged value for each layer as a function of the normalized strain is shown in the main part of the figure. We find that with increasing strain, the highly folded configurations indicated by $t_4 \approx \pi$ extend to larger number of collapsed layers $n$.
  
To determine the theoretical strain value associated with successive snapping of layers onto the inner cylinder, we simulate an isotropic deformation of the same origami geometry (Methods). We plot the number of collapsed layers as a function of the applied normalized strain (black line and open circles). We find that these data track the experimentally determined interface between the open versus collapsed vertex structures denoted by high values of $t_4$. Furthermore, we find that comparison with $dF/d\gamma_x$ (vertical dashed lines) identifies where the structure undergoes an abrupt drop in compressive modulus due to each sequential snap (labeled by $n$). 

Collectively, these studies demonstrate a powerful approach for designing and programing specific mechanical properties into origami systems. The structures we considered here are degree-6 Miura-ori vertices. Our methodology however, is general. Thus, the application of modular designs and propagation of constraints between vertices to origami patterns with different DOF will undoubtedly shed light on other nontrivial mechanical mechanisms and design principles.  Moreover, these results naturally extend to other systems. For example, the interactions between vertex structures described here is reminiscent of set-valued maps commonly adopted in studies of robotic, biological, physical, and economic systems \cite{Aubin:1990vqa}. There, the evolving dynamics produce constrained states that correspond to the allowed configurations we discussed. This set of constrained states, in turn, determines the possible evolution pathways for the system to achieve its next set of viable states. It may be possible to employ the same modular and bifurcating designs described here to these other systems in order to better control their behaviors. The advantage of this approach being that it introduces a ``robustness'' protected by the configuration space topology under different constraints.

\section{Method}
\subsection{Three-dimensional model of origami structures}
A mathematical model of each experimentally generated origami structure is formulated for numerical simulation and 3D reconstruction. The model is composed of the vertex coordinates and the constraints imposed by the crease length between neighboring vertices. Given a 2D projection of all the vertex positions ($x_i$, $y_i$), the $z_i$ coordinates in the third dimension are obtained by minimizing a penalty function \cite{Tachi:2013im} $V(z)=\frac{k}{2}  \sum_{i,j} ' \left[\sqrt{(x_i-x_j)^2+(y_i-y_j)^2+(z_i-z_j)^2} - l_{ij}\right] ^2$, where $l_{ij}$ is the length of the crease between two connected vertices of indices $i$ and $j$ on the triangular crease lattice. The $z$ coordinates can thus be obtained by solving the associated ordinary differential equation (ODE) array as $\dot{z_i}=-k \sum_j '  \left[\sqrt{(x_i-x_j)^2+(y_i-y_j)^2+(z_i-z_j)^2} - l_{ij}\right] \hat{r}_{ij} \cdot \hat{z}$, where the unit vector $\hat{r}_{ij}=((x_i-x_j)\hat{x}+(y_i-y_j)\hat{y}+(z_i-z_j)\hat{z})/\sqrt{(x_i-x_j)^2+(y_i-y_j)^2+(z_i-z_j)^2}$, and $k$ is the virtual spring constant for perturbing the crease lengths. The ambiguity due to mountain-valley alignments is avoided by shifting the vertices up or down with a small amount in the $z$ direction as the initial condition. 

\subsection{Configuration space analysis}
The configuration space of each vertex is described by the accessible folding angles of the connected creases. In the case of a degree-6 vertex, the configuration space is three dimensional (since DOF=3). Any potentially foldable structure is realized by rotating the creases through those 3 independent folding angles. The foldability is validated by the criterion that the distance between any pair of transformed vertices cannot be greater than that in the flat unfolded state for any inextensible sheet. Self-interceptions of the structures are also identified and excluded from the configuration space (see SI).

\subsection{Experiments}

The origami structures were made from a flat paper sheet (Stardream Metallics 81lb) with measured thickness $\tau=0.16$ mm and flexural rigidity $D=4.0 \times 10^{-4}$ N$\cdot$m \cite{Silverberg:2015gb}. The crease lines were perforated by a laser cutter and weakened manually by flat-folding along the perforation lines in both directions. Its folding kinematics were recorded by a top-view camera, which was calibrated for extracting the vertex locations. The $x-y$ coordinates of the vertices were fed into the aforementioned mathematical model of the origami to extract the $z$ depths of the vertices and the 3D configurations of the folded origami sheets were reconstructed using Matlab. 

\section{Acknowledgments}
\begin{acknowledgments}
The authors thank Andy Ruina, Tim Healy, James Jenkins, Uyen Nguyen, Lea Freni, and the Cohen lab for useful discussions.  
We also thank F. Parish for assistance with the laser cutter. This work was supported by the National Science Foundation Grant No.~EFRI ODISSEI-1240441. I.C. received continuing support from DMREF-1435829. C.D.S. thanks the KITP, Santa Barbara, for hospitality. 
\end{acknowledgments}

\newpage

\end{document}